\documentclass[useAMS,usenatbib]{mn2e}
\usepackage{aas_macros,graphicx,times,multirow}

\newcommand{\vltn}{{Very Large Telescope}}
\newcommand{\vlt}{{VLT}}
\newcommand{\hstn}{{\em  Hubble  Space Telescope}}
\newcommand{\hst}{{\em HST}}

\newcommand{\xmm}{{\em XMM-Newton}}

\def \rxj{RX\, J1856.5$-$3754}

\title[The birthplace and age of the isolated neutron star RX\, J1856.5$-$3754]{The birthplace and age of  RX\, J1856.5$-$3754}
\author[R. P. Mignani, et al.]
{\parbox{\textwidth}{R. P. Mignani$^{1,2}$\thanks{E-mail:rm2@mssl.ucl.ac.uk}, 
D. Vande Putte$^{1}$, 
M. Cropper$^{1}$, 
R. Turolla$^{3,1}$, 
S. Zane$^{1}$, 
L. J. Pellizza$^{4}$, 
L. A. Bignone$^{4}$, 
N. Sartore$^{5}$, 
A. Treves$^{6}$} \\ \\
$^{1}$ Mullard Space Science Laboratory, University College London, Holmbury St. Mary, Dorking, Surrey, RH5 6NT, UK\\
$^{2}$ Kepler Institute of Astronomy, University of Zielona G\'ora, Lubuska 2, 65-265, Zielona G\'ora, Poland \\
$^{3}$ Dipartimento di Fisica e Astronomia, Universit\'a di Padova, via Marzolo 8, 35131 Padova, Italy \\
$^{4}$ Instituto de Astronom\'ia y F\'isica del Espacio, UBAÐCONICET, C.C. 67, Suc. 28 (C1428ZAA),Buenos Aires, Argentina \\
$^{5}$ INAF - Istituto di Astrofisica Spaziale e Fisica Cosmica Milano, via E. Bassini 15, 20133, Milano, Italy \\
$^{6}$ Dipartimento di Fisica e Matematica, Universit\'a dell'Insubria, via Valleggio 11, 22100, Como, Italy \\ 
 }
\begin{document}

\date{Accepted 1988 December 15. Received 1988 December 14; in original form 1988 October 11}

\pagerange{\pageref{firstpage}--\pageref{lastpage}} \pubyear{2002}

\maketitle

\label{firstpage}

\begin{abstract}
X-ray observations unveiled various types of radio-silent Isolated Neutron Stars (INSs), phenomenologically very diverse, e.g. the $\sim$Myr old X-ray Dim  INS (XDINSs) and the $\sim$kyr old magnetars. Although their phenomenology is much diverse, the similar periods ($P$=2--10 s) and magnetic fields ($\approx10^{14}$ G) suggest that XDINSs are evolved magnetars, possibly born from similar populations of supermassive stars. One way to test this hypothesis is to identify their parental star clusters by extrapolating backward the neutron star velocity vector in the Galactic potential. By using the information on the age and space velocity of the XDINS \rxj, we computed backwards its orbit in the Galactic potential and searched for its parental stellar cluster by means of a  closest approach criterion. We found a very likely association with the Upper Scorpius OB association, for a neutron star age of $0.42\pm0.08$ Myr, a radial velocity $V_r^{\rm NS} =67\pm 13$ km s$^{-1}$, and a present-time parallactic distance $d_{\pi}^{\rm NS} = 123^{+11}_{-15}$ pc. Our result confirms that the ``true'' neutron star age is much lower than the spin-down age ($\tau_{sd}=$3.8 Myrs), and is in good agreement with the cooling age, as computed within standard cooling scenarios. The mismatch between the spin-down and the dynamical/cooling age would require either an anomalously large breaking index ($n\sim 20$) or a decaying magnetic field with initial value $B_0\approx 10^{14}$ G. Unfortunately, owing to the uncertainty on the age of the Upper Scorpius OB association and the masses of its members we cannot yet draw firm conclusions on the estimated mass of the \rxj\ progenitor.
\end{abstract}


\section{Introduction}

X-ray  observations  performed in the last two decades unveiled the  existence  of Isolated Neutron Stars (INSs) which are mostly radio-silent and not  powered by the star rotation, such as the Myr-old X-ray Dim INS, or XDINSs (Turolla\ 2009), and the kyr-old magnetar  candidates (Mereghetti\ 2008). Despite their different phenomenology,  with XDINSs featuring stable and thermal soft X-ray emission ($kT \sim$ 50--100  eV) and magnetars featuring transient and bursting high-energy activity and non-thermal spectral tails,  both INS classes  are thought to be linked by evolution. This is implied by their close locations in the period-period derivative ($P$-$\dot{P}$) diagram and their similar values of the surface magnetic fields $B_{surf}$, inferred either from the neutron star spin down or by the observation of absorption features in the X-ray spectra, that can be both in the $\approx 10^{13}$--$10^{14}$ G range. 
While  finding more similarities in their multi-wavelength emission  would strengthen such a  link, confirming it is more challenging. 

If linked by evolution, both magnetars and XDINSs should follow a  common evolutionary path. Interestingly, some magnetars  seem to be associated  with super-massive star clusters (Muno et al.\  2005),  thus with putative progenitors of $\ga 40 M_\odot$, more massive than those of ``normal'' neutron stars (8--25 $M_\odot$; Heger et al.\  2003). However,  identifying the parental clusters of the XDINSs is more complicated because of their larger ages with respect to the magnetars. Indeed, the results are affected by the uncertainty on the orbit extrapolation in the  Galactic potential for time scales of a few Myrs. This depends on unknowns like the neutron star distance and proper motion (hence its tangential velocity), measured from optical astrometry once the counterpart is known, and the neutron star age, inferred either from the spin-down ($\tau_{sd}$), once the period derivative is measured, or from the neutron star cooling ($\tau_{cool}$), once both a model and a reference value of the surface temperature are assumed. A further unknown is the neutron star radial velocity, whose uncertainty dramatically increases the chances of spurious matches with clusters or  stellar associations and hampers all identification attempts (e.g., Mignani et al.\ 2010). However, in some cases the radial velocity can be inferred from the modelling of the bow-shock profile formed as the INS moves supersonically through the interstellar medium (ISM) and its fitted inclination angle with respect to the line of sight (LOS), like it had been done for Geminga (Pellizza et al.\ 2005).The association with the parental cluster also gives an estimate of the neutron star dynamical age ($\tau_{dyn}$).

The XDINS RX\, J1856.4$-$3754 (Walter et al.\ 1996) is the best target for this goal.  It has the brightest optical counterpart amongst XDINSs ($V\sim25.7$; Walter \& Matthews 1997) and its proper motion and parallactic distance have been measured with high accuracy with the \hstn\ (\hst), e.g. Walter et al.\ (2010), while the radial velocity has been inferred by modelling the bow-shock (van Kerkwijk \& Kulkarni\ 2001a)   detected in H$\alpha$ by the \vltn\  (\vlt). Moreover, the period derivative of \rxj\ has been measured (van Kerkwijk \& Kaplan 2008) yielding the value of $\tau_{sd}=$3.8 Myrs. 
In this paper, we report on the search for the parental stellar association of \rxj, based on the backward extrapolation of its orbit in the Galactic potential. The paper is divided as follows: the description of the orbit simulation code and its application to to the neutron star and to candidate parental clusters/OB associations is given in Sectn.\ 2, the results are presented in Sectn.\ 3 and discussed in Sectn.\ 4, respectively. 

\section{Orbit simulation}

For the  Galactic orbit simulation we  used the code of Vande Putte \& Cropper  (2009), already successfully applied in Rauch et al.\ (2007) and in Vande Putte et al.\ (2009; 2010) and we refer to these publications for further details. As discussed in Sectn.\ 1, in order to extrapolate the orbit of \rxj\ in the  Galactic potential and localise its putative birth place, accurate measurements of its proper motion $\mu^{\rm NS}$, parallactic distance $d_{\pi}^{\rm NS}$, and inclination angle $i$ of the space velocity vector along the LOS are required. Both the proper motion and parallactic distance of \rxj\ have been repeatedly measured through optical astrometry techniques with the \hst. As seen from Table \ref{astro},  all measurements of the \rxj\ proper motion agree within the quoted uncertainties.  For the parallactic distance, we assumed  the most recent value of Walter et al.\ (2010), which confirms the earlier measurement of Walter \& Lattimer (2002) and is consistent with that of Kaplan et al.\ (2002), but it is more accurate.\footnote{The first measurement of the parallactic distance, $d_{\pi}^{\rm NS} =61^{+9}_{-8}$ (Walter 2001), was not confirmed by Walter \& Lattimer (2002) and  Kaplan et al.\ (2002) and is not listed in Table 1, while those quoted in van Kerkwijk \& Kaplan (2007) and Kaplan et al.\  (2007) were presented without any supported evidence and are listed for completeness only.}
The inclination angle $i$ along the LOS has been measured by van Kerkwijk \& Kulkarni\  (2001a) by fitting the intensity profile of the bow-shock detected in H$\alpha$ by the \vlt. This is $i=60^{\circ} \pm 15^{\circ}$, which means that the neutron star would not move far from the plane of the sky. The alternative ionisation nebula model considered by van Kerkwijk \& Kulkarni (2001a), which yielded inclination angles closer to the LOS,  has been ruled out by Kaplan et al.\ (2002) because the nebula's opening angle would be incompatible with any of the parallactic distance measurements.

\begin{table}
\begin{center}
  \caption{Compilation of the proper motion ($\mu^{\rm NS}$) and parallactic distance ($d_{\pi}^{\rm NS}$) measurements for \rxj.}
  \label{astro}
\begin{tabular}{cll} \\ \hline
Parameter & value & Reference   \\  \hline $\mu^{\rm NS}$
                        & $332 \pm 1$     & Walter\ (2001) \\
                       (mas yr$^{-1}$)  & $333 \pm 1$     & Kaplan et al.\ (2002) \\
                        & $331.2\pm 2.0$  &  Walter et al.\  (2010) \\ \hline
$d_{\pi}^{\rm NS}$                             & $140 \pm 40$   & Kaplan et al.\ (2002) \\
 (pc)                       & $117 \pm 12$  & Walter \& Lattimer (2002) \\
                        & $161^{+18}_{-14}$ & van Kerkwijk \& Kaplan (2007)$^{*}$ \\
                        & $167^{+18}_{-15}$  &  Kaplan et al.\  (2007)$^{*}$ \\
                        & $123^{+11}_{-15}$   &  Walter et al.\ (2010) \\ \hline
\end{tabular}
\label{data}
\end{center}
$^{*}$ This value has not been used in the current analysis
\end{table}

Using our orbit simulation code, we then calculated several test sets of neutron star tracks by looping on various values of the neutron star distance, the inclination angle (hence of the tangential and radial velocity), and the neutron star age, computed around their reference values. In particular, we considered a grid of distance values  which are $\pm 3 \sigma$ around the best fit value of Walter et al.\ (2010), sampled with a spacing of 10 pc.  From the corresponding sampled values of the transverse velocity, we then computed a grid of values for the radial velocity for different values of the inclination angle $i$ (45$^{\circ}$--75$^{\circ}$) sampled with a $5^{\circ}$ step. For the backward extrapolation time, we considered a grid of values for the neutron star age sampled at intervals of 10 kyrs. Since the uncertainty on the \rxj\ proper motion is $\la 0.6\%$, we neglected its influence on the orbit extrapolation and the neutron star birth place localisation.  In the orbit computation, we did not account for possible changes in the neutron star spatial velocity caused by close encounters with other stars or known star clusters.

As a reference for the search of the parental stellar association, we firstly considered a sample of open clusters (OCs) selected from the latest version  (3.1, released on 2010 November 24) of the data compilation of Dias et al.\ (2002), also referred to as the DAML catalogue. This contains entries for  2140 OCs, with sky coordinates, proper motion, radial velocity, and associated errors, together with information on the cluster metallicity, size, colour excess, Trumpler type, and age.   Distances in the DAML catalogue are reported without associated errors, although a fiducial 10\% uncertainty is probably adequate for most cases (see, e.g. Vande Putte et al.\ 2010).  We also used as a reference the catalogues of nearby OB associations of de Zeeuw et al.\ (1999) and Mel'nik \& Dambis \ (2009), based on {\em Hipparcos} data.  Both catalogues contain sky positions and proper motions in Galactic coordinates, radial velocities, and errors. In the de Zeeuw et al.\  (1999) catalogue, radial velocities are reported with no associated errors and we assumed a fiducial 20\% uncertainty. Distances are derived from the trigonometric parallaxes of association star members (if available) or from photometric parallaxes. For the OB associations in  the Mel'nik \& Dambis \ (2009) catalogue we assumed their quoted 6\% uncertainty on the distance, while for both proper motion and radial velocity (for which no error is given) we assumed a conservative 50\% uncertainty. In all cases, we selected entries with non-null values of distance, proper motion, radial velocity (and associated errors). Although objects younger than  $\sim100$ Myr are obviously the most interesting  candidates, initially we did not apply any selection based on the age  of the OC or OB association, which can be uncertain by up to $\sim 40\%$, as well as on other parameters, like the metallicity or morphological type.  Instead, we decided to use these parameters as validation elements once a potential association was found. Our list include  439 OCs from the DAML catalogue and  77 OB associations from the de Zeeuw et al.\ (1999) and Me\l'nik \& Dambis \ (2009) catalogues.

\begin{table*}
\begin{center}
  \caption{Name, coordinates, right ascension and declination  proper motions ($\mu_{\alpha}$; $\mu_{\delta}$), distances $d$, and  radial velocities $V_{r}$  for the OC and OB associations in the  Dias et al.\ (2002) and de Zeeuw et al.\  (1999) catalogues which are potential birth places for \rxj\  (See Sectn.\ 3).   Values of the closest approach $\Delta r$ and the corresponding neutron star dynamical age ($\tau_{dyn}$), distance ($d^{\rm NS}$), and radial velocity ($V_r^{\rm NS}$) are given in the last four columns.}
  \label{ob}
\begin{tabular}{lcccccc|cccc} \\ \hline
Name & RA & Dec & $\mu_{\alpha}$ & $\mu_{\delta}$ & d & $V_{r}$  &  $\Delta r$ & $\tau_{dyn}$ & $d^{\rm NS}$  & $V_r^{\rm NS}$ \\ 
          & $^{(hms)}$  & $(^{\circ} ~ \arcmin ~ \arcsec)$ & (mas yr$^{-1}$) & (mas yr$^{-1}$) & (pc) & (km s$^{-1}$) & (pc) & (Myr) & (pc) & (km s$^{-1}$) \\ \hline
NGC\,6475      & 17 53 51  & -34 47 36  &  +1.67 $\pm$  0.20  &  -3.60  $\pm$  0.20  &  301   &  -15.53  $\pm$  1.04 & 160--170 &  0.1--0.3 & 140--150 &  60--180\\
ASCC\,99       & 18 49 50  & -18 43 48  &  +6.90 $\pm$  0.64  &  -2.50  $\pm$  0.49  &  280   &  -31.29  $\pm$  0.40 & 150--160 & 0.1--0.2  & 140--150 & 60--210 \\ \hline
Up-Sco             & 16 12 03 & -23 25 09 & +11.04$\pm$0.01 & -23.32$\pm$0.14 & 145$\pm$2  & -4.6 $\pm$  0.92& 5--25 & 0.3--0.5 & 120--150 & 50--150  \\
Up-CenLupus   & 15 08 12 & -43 45 06 & +21.30$\pm$0.35 & -23.13$\pm$0.14 & 140$\pm$2  & +4.9  $\pm$ 0.98 & 55 & 0.3--0.5 & 110--150 & 50--180 \\
Lo-CenCrux      & 12 18 52 & -57 05 29 & +33.50$\pm$0.11 & -8.90$\pm$0.09  & 118$\pm$2  & +12.0 $\pm$  2.4 & 95 & 0.3--0.4 & 90--110 & 90--190 \\  \hline
\label{obs}
\end{tabular}
\end{center}
\end{table*}

We  extrapolated back in time the orbits of the candidate parental OC and OB associations over the same age range as for \rxj, using as a reference their nominal values of distance, proper motion, and radial velocity. Then, we looked for the combination of neutron star parameters (age, distance, radial velocity) which yielded the closest approach of the \rxj\ orbit. We then regarded as likely associations those for which the approach was closer than a given threshold, defined as the overall uncertainty on the computed separation. The association threshold accounts for the uncertainty on the cluster/OB association orbit extrapolation due to the random errors associated with their distance, proper motion, and radial velocity. As done in Vande Putte et al.\ (2010), we estimated this uncertainty through a Monte Carlo simulation. For each object, we simulated 1000  different values of the distance, proper motion, and radial velocity, sampled within their formal errors, and computed the root mean square (rms) of the separation between their backward-extrapolated positions at the reference and the centre of the Galactocentric reference frame.  In most cases, we found that the uncertainty on the orbit extrapolation was below 100 pc.   The association threshold also accounts for the spatial extent of the OC or OB association computed from its angular size and distance and assuming, as a first approximation, a spherical symmetry. Since most OCs and OB associations tend to have irregular morphologies, this is the most conservative assumption we can make.  For simplicity, we did not account for two opposite effects which could influence the actual OC or OB association angular size in the past: its expansion due to intrinsic member star proper motions and radial velocities, and its evaporation due to star escape from the local gravitational potential, which would yield angular sizes smaller and larger that those measured at the present epoch, respectively.

\section{Results}

For completeness, we initially explored an age range of 2.8--3.8 Myr around the \rxj\ spin-down age, with the caveat that this is an intrinsically uncertain age indicator since it depends on both the  initial spin period of the neutron star and the value of the braking index $n$, which has not been measured yet for \rxj.  For  the assumed range of parameters we could not find a  likely cluster association for \rxj.   Although, it is possible that its parental OC/OB association is not a  known one, owing to the relatively small distance ($\approx 1.5$--2.5 kpc) travelled in 2.8--4.8 Myr it is unlikely that it has not been discovered yet.  While it is also possible that it has been filtered out in the sample selection (see Sectn.\ 2), the most likely conclusion is that the explored age range is not representative of the \rxj\ age.

This conclusions is confirmed by our measurement of its cooling age $\tau_{cool}$, computed using as a reference the most recent measurements of its surface temperature obtained from X-ray and optical-UV  observations.  It has been previously suggested that the surface temperature  of \rxj\ is non-uniform (Pons et al.\ 2002; Braje \& Romani 2002; Tru\"mper et al.\ 2004), as in other cooling INSs. The discovery of X-ray pulsations at a period of $\sim 7$ s (Tiengo \& Mereghetti 2007) supports this picture.  Indeed, in a recent analysis of archival \xmm\ observations of \rxj, covering a time span of almost 10 years, Sartore et al.\ (2012) found that a two blackbody (BB) model is statistically favoured with respect to a single BB in order to describe the 0.15--1.2 keV spectrum. The resulting BB temperatures are $T_h$ = 62.4 eV and $T_c$ = 38.9 eV  for the hot and cold components, while the corresponding BB radii are $R_h$ = 4.7 (d/120 pc) km and $R_c$ = 11.8 (d/120 pc) km, respectively. When extrapolated to optical-UV wavelengths, the combined emission of the two BBs is consistent with the optical-UV fluxes obtained from  {\em HST} photometry (Kaplan et al.\ 2011), and further supports the two-BB picture. This gives a luminosity (at infinity) of  $31.4 \leq \log L\,{\rm (erg\, s^{-1})}\leq 32.5$, where the uncertainty is computed accounting both for the unknown viewing geometry and the uncertainty on the distance (Walter et al.\ 2010).   This range of luminosity values is fully consistent with that of the single BB, $31.2 \leq \log L\,{\rm (erg\, s^{-1})}\leq 31.9$, and those reported in the literature (see, e.g. Burwitz et al.\ 2003; van Kerkwijk \& Kaplan 2007; Walter \& Lattimer 2010) and lead to the same conclusions on the source age. Assuming a minimal cooling scenario (Page et al.\ 2006, 2009), such luminosities imply cooling ages of  $\approx$ 0.1--1 Myr, depending on the star mass and chemical composition.  This is incompatible with the estimated spin-down age ($\sim 4\, {\rm Myr}$) of \rxj, unless the neutron star is closer than 90 pc, which is only marginally consistent with the $3 \sigma$ uncertainty on the parallactic distance (Walter et al.\ 2010).   To summarise, the estimates of the cooling age $\tau_{cool}$ suggest neutron star  ages as low as $\sim$ 0.1 Myr.  Thus, we reran our orbit simulations exploring the dynamical age range  $0.1$--2.8 Myr.

We found that the closest OC associations are with NGC\, 6475 and ASCC\,99 (see Table  2). For the former, the closest separation $\Delta r= 160$--170  pc for a neutron star  age $\tau_{dyn}=0.1$--0.3 Myr, a present-time distance $d^{\rm NS}=140$--150 pc, and a radial velocity $V_r^{\rm NS} =60$--180 km s$^{-1}$, while for the latter the closest separation $\Delta r= 150$--160 for  $\tau_{dyn}=0.1$--0.2 Myr, $d^{\rm NS}=140$--150 pc, and $V_r^{\rm NS} =60$--210 km s$^{-1}$. In both cases, the values of the closest separations are well above their corresponding association threshold ($\sim 15$ pc).  Moreover, the estimated OC ages of $\sim 0.3$ and $\sim 0.5$ Gyr  (Dias et al.\  2002) would be much larger than the inferred neutron star dynamical age $\tau_{dyn}$ of 0.1--0.3 Myr and that of its putative massive progenitor (30--60 Myr). Thus, we ruled out the possible associations with NGC\, 6475 and ASCC\,99.
Similarly, we found possible associations with the Upper Scorpius, Upper Cen  Lupus, and Lower Cen Crux OB associations. For the Upper Cen  Lupus, the closest approach separation $\Delta r \sim 55$ pc, for  $\tau_{dyn} =0.3$--0.5 Myr,  $d^{\rm NS}=110$--150 pc, and   $V_r^{\rm NS} = 50$--180 km s$^{-1}$, while for the Lower Cen Crux we derived  $\Delta r \sim 95$ pc  for $\tau_{dyn}=0.3$--0.4 Myr, $d^{\rm NS}=90$--110 pc, and $V_r^{\rm NS} =90$--190 km s$^{-1}$. However, only for the Upper Scorpius  the closest approach distance is below the corresponding association threshold ($\sim$ 25 pc). We note that this value is dominated by the size of the OB association (de Zeeuw et al.\ 1999) rather than by the uncertainty on its orbit extrapolation, which is quite low thanks to the small uncertainties on the proper motion, distance, and radial velocity derived by {\em Hipparcos} (see Table 2).
In particular, for the Upper Scorpius we obtained  $\Delta r= 5$--25  pc for $\tau_{dyn}=0.3$--0.5 Myr,  $d^{\rm NS}=120$--150 pc, and $V_r^{\rm NS} =50$--150 km s$^{-1}$. The inferred range of values for the neutron star present-time distance is consistent with that derived from the parallax measurement of Walter et al.\ (2010), $d_{\pi}^{\rm NS}=123^{+11}_{-15}$ pc.
Conversely, fixing the neutron star  present-time distance to this value would yield $\tau_{dyn} =0.42 \pm 0.08$ Myr,  $V_r^{\rm NS} = 67 \pm 13$ km s$^{-1}$,  and a closest approach separation $\Delta r = 17 \pm 8$ pc, after accounting for the uncertainty on the parallactic distance. Thus, we deem the Upper Scorpius OB association as the putative parental cluster for \rxj.

\section{Discussion and conclusions}

An association with the Upper Scorpius was originally proposed, e.g.  also by Kaplan et al.\ (2002) and Walter \& Lattimer (2002), assuming their own measurements of the \rxj\ parallactic distance and leaving its radial velocity unconstrained. Their inferred dynamical ages for \rxj\ were $\tau_{dyn}\sim 0.4$ Myr and $\sim 0.5$ Myr,  respectively, close to the values inferred in the current work.
Our conclusion is also in line with the more recent results of Tetzlaff et al.\ (2010; 2011). In particular, for the same value of the neutron star distance $d_{\pi}^{\rm NS}=123^{+11}_{-15}$ pc (Walter et al.\ 2010) as assumed by Tetzlaff et al.\ (2011), the dynamical age that we derive for the neutron star ($\tau_{dyn} =0.42 \pm 0.08$ Myr) is fully consistent with their estimate ($0.46 \pm 0.05$ Myr).
The neutron star radial velocity that we infer ($67 \pm 13$ km s$^{-1}$), though, is larger than that derived by Tetzlaff et al.\ (2011), $6^{+19}_{-20}$ km s$^{-1}$, with any effect due the difference in the assumed proper motion value being negligible. The difference in the inferred radial velocity (still below $3\sigma$) is probably ascribed to the different approach in selecting parameter values for the orbit simulation. Tetzlaff et al.\ (2011) sampled a uniform radial velocity distribution in the range -250--250 km s$^{-1}$, to consider also directions much closer to the LOS, as predicted by the ionisation nebula model. Instead, by assuming the more likely bow-shock model (see footnote 2), we only sampled values in the range 50--200 km s$^{-1}$, according to the constraints on the bow-shock inclination angle along the LOS.  The difference in the inferred radial velocity can, thus, explain a factor of 2 difference in the closest approach separation (for the same value of the neutron star distance).
A similar result as ours has been obtained by Bignone et~al. (in prep.) using a different orbit simulation code and a Monte Carlo approach to estimate the probability of a parental cluster association. They also find that Upper Scorpius is the only candidate with a non-negligible probability of being the \rxj\ parent system and obtain a value of $\tau_{\rm dyn} \sim 0.3\ {\rm Myr}$, compatible with ours.
To summarise,  the agreement between these three  results rules out that the  proposed association between \rxj\ and the Upper Scorpius  is affected by systematic effects and makes it statistically robust.

Identifying the Upper Scorpius OB association as the birthplace of \rxj\ yields a  robust value of the neutron star dynamical age $\tau_{dyn} = 0.42\pm 0.08$ Myr, which is almost an order of magnitude lower than the spin-down age $\tau_{sd} = 3.8$ Myr. The dynamical age we derived is  well compatible with the age, $\tau_{cool}$, required to reproduce the observed  source luminosity in standard cooling models (the minimal cooling scenario, Page et al.\ 2006, 2009). ``Fast'' (or enhanced) cooling (e.g., Yakovlev, Levenfish  \& Shibanov 1999 for a review) is not required to explain the observed properties of  \rxj.
The mismatch between the spin-down and the dynamical/cooling age implies that  the characteristic braking  index of the source is $n\sim 2\tau_{sd}/\tau_{dyn}\approx 20$. Large and positive braking indices are routinely measured in radio-pulsars (e.g. Johnston \& Galloway 1999).  However, assuming that such a large $n$ represents the average (constant in time) value in  the standard expression for spin-down $\dot\nu =-k\nu^n$, where $\nu$ is the star spin frequency and $k$ a constant, would rule out any known form of braking torques. A  possibility is to consider standard spin-down by magneto-dipole losses ($n=3$ in the previous expression) but in the presence of a decaying magnetic field, which results in $k\propto B^2$ decreasing in time.
Self-consistent models for the coupled magnetic and thermal evolution of a  neutron star were recently presented by Pons, Miralles \& Geppert (2009), to which we refer for details. Starting from a series of models with different mass $M$ and initial surface dipolar field $B_0$, we looked for values of $M$ and $B_0$ which reproduce the present value of the luminosity $L$ (two representative values $5  \times 10^{31}$  and $10^{32}$ erg s$^{-1}$ were chosen) at an age  equal to the dynamical age, 0.42 Myr. This is
achieved by models with $B_0\approx 10^{14}$ G and $M$ in the range 1.1--1.7 $M_\odot$. Further imposing that the period matches the  observed one, $P\simeq 7$ s (Tiengo \& Mereghetti 2007)  restricts the mass to be $\ga 1.3\,  M_\odot$. This further supports our conclusion that the true age of \rxj\ is sensibly shorter than  the spin-down age and well compatible with our inferred dynamical age. The lower limit on the star mass when combined with the estimate of the star radius obtained by Sartore et al.\ (2012) from the two-blackbody fit, $12.5 \la R\, \rm{(km)}\la 17.3$, is rather non-constraining for the neutron star equation of state; we remark also that these estimates of $M$ and $R$ are model-dependent and should be taken with caution.
A further possibility to reconcile the dynamical and spin-down ages is that  \rxj\ was  born with an uncommonly long period. Using the standard spin-down formula ($P^2-P_0^2\propto B^2 t$), it easy to verify that to spin down the star at its present 7 s period in a time $t = \tau_{dyn}=0.42\, {\rm Myr}$ an initial period $P_0\sim 6.5\, {\rm s}$ is required for $B\sim 1.5\times 10^{13}\, {\rm G}$.

If, as it seems likely, \rxj\ originated $\sim 0.4\, {\rm Myr}$ ago in the Upper Scorpius OB association, we can constrain the mass of the progenitor.  The stellar population in Upper Scorpius has been extensively studied in the past (e.g. Priebisch et al.\  2002 and references therein). The estimated age of the association is $\approx$ 5 Myr, according to mass estimates of Antares (an evolved M1 Ib supergiant, $\approx 20$--25$M_\odot$) and of the runaway OB star $\zeta$ Oph ($\approx 20 M_\odot$). Hence, the putative progenitor of \rxj, which went SN $\approx$ 0.4 Myr ago, must have had a mass of 20--60  $M_\odot$, larger than usually expected for neutron star progenitors ($8 M_\odot \la M \la 25 M_\odot$; Heger et al.\ 2003) and close to those invoked  for magnetar progenitors  (Muno et al.\ 2005). This, together with the high initial magnetic field required to reconcile the cooling and dynamical age in the framework of a decaying magnetic field, would then suggest that \rxj\ might have been an active magnetar in its youth. However, a recent re-assessment of the age of the Upper Scorpius association based on optical spectroscopy (Pecaut et al.\ 2012) suggests a more likely age of $\approx$ 11 Myr. Hence, the masses of all members of the association would be lower than previously estimated (e.g. the mass of Antares would be $\approx 17 M_\odot$) and the mass of the \rxj\ progenitor would be $\approx 18$--20 $M_\odot$, i.e. still in the range expected for "normal" neutron star progenitors. {\em If} one assumes that magnetars are indeed born from supermassive progenitors, this might suggest that \rxj\ is not an evolved magnetar. However, any conclusion has to be taken with due care.  First of all, the estimates on the Upper Scorpius age and on the mass of its members are still debated, which makes it difficult to precisely constrain the mass of the \rxj\ progenitor.  Second of all, any conclusion on the XDINS progenitors' masses must be verified against other XDINS/cluster associations.  The only other XDINS for which both the optical parallax and proper motions have been measured (Kaplan et al.\ 2007), and for which the Galactic orbit extrapolation can be reasonably well computed, is RX\ J0720.4$-$3125.  The source has been associated with the Trumpler 10 OB association (Kaplan et al.\ 2007; Tetzlaff et al.\ 2011) for a dynamical age of 0.5--0.8 Myr. This would imply a progenitor no more massive than $\approx 20 M_\odot$, again close to that expected for "normal" neutron star progenitors.  However, owing to the lack of information on the neutron star radial velocity from a yet undetected bow-shock, the association with Trumpler 10 is still tentative. The other way round, {\em if} one assumes that XDINSs are indeed evolved magnetars, the possible association of \rxj\ with a relatively low-mass progenitor might suggest that not all magnetars are born from super-massive stars.  Indeed, the proper motions of the magnetars  XTE\, J1810$-$197 (Helfand et al.\ 2007), PSR\, J1550$-$5418 (Deller et al.\ 2012), SGR\, 1806$-$20, and SGR\, 1900+14 (Tendulkar et al.\ 2012), imply transverse velocities $\la 300$ km s$^{-1}$, which would not require hyper-energetic SN explosions from super-massive progenitors. 

To summarise, although it seems likely that the XDINS \rxj\ was born $\sim 0.4$ Myr ago in the Upper Scorpius OB association, both the uncertainty on the age of the latter and the masses of its members make it difficult to draw firm conclusions on the actual mass of the neutron star progenitor, hence verify the possible evolutionary link between XDINSs and magnetars. Establishing parental cluster associations for a larger sample of XIDNSs and magnetars, together with accurate studies of the cluster stellar population, is crucial to verify the connection between the properties of their progenitors and the evolutionary tracks followed by the neutron star after the SN explosion.

\section*{Acknowledgments}
We thank J. Pons for  providing us with a set of his cooling models, T. de Zeeuw for useful discussions, and the anonymous referee, whose comments contributed to improve our manuscript. NS acknowledges the support of ASI/INAF through grant I/009/10/0.

\label{lastpage}

\end{document}